\documentclass[aps,prl,reprint,groupedaddress]{revtex4-1}

\usepackage[dvipdfmx]{graphicx}
\usepackage{latexsym}
\usepackage[fleqn]{amsmath}
\usepackage{amssymb}

\begin{document}

\title{Solvable Chaotic Synchronization \\--A New Interpretation of Common
Noise-induced Synchronization \\ with Conditional Lyapunov Exponents--}


\author{Masaru Shintani}
\email[shintani.masaru.28a@st.kyoto-u.ac.jp]{}
\author{Ken Umeno}%
\email[umeno.ken.8z@kyoto-u.ac.jp]{}
\affiliation{Department of Applied Mathematics and Physics, Graduate School of Informatics, Kyoto University,Yoshida Honmachi Sakyo-ku,Kyoto, 606--8501}


\date{\today}

\begin{abstract}
 We present a solvable chaotic synchronization model of
 unidirectionally coupled dynamical systems. We establish a
 new interpretation of the conditional Lyapunov exponent that
 characterizes chaotic synchronization completely. Moreover, we newly show how the conditional Lyapunov
exponent relates to common noise-induced synchronization phenomena by the new interpretation.
\end{abstract}
\pacs{4}

\maketitle

\begin{figure}[ht]
  \begin{center}
    \includegraphics[width=80mm]{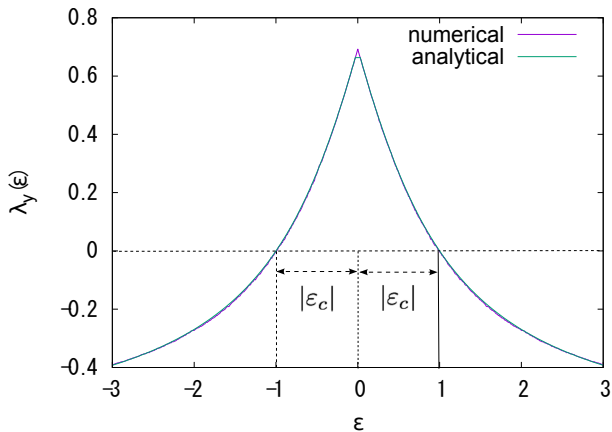}
     \caption{Conditional Lyapunov exponent  $\lambda_{y}(\varepsilon)$}
   \label{fig:compare_cle}
  \end{center}
\end{figure}
\begin{figure}[tb]
  \begin{center}
    \includegraphics[width=80mm]{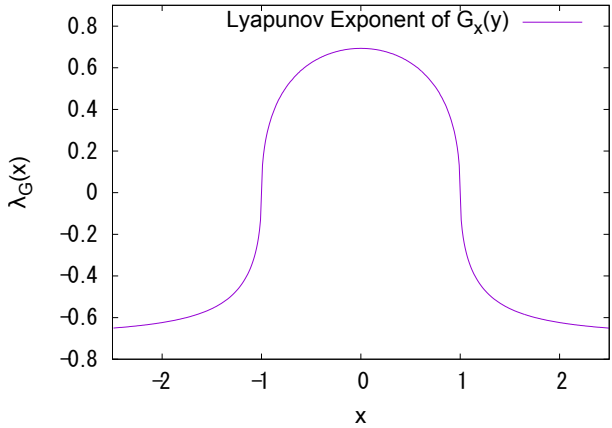}
    \begin{center}
     \caption{the relation between $x$ and $\lambda_G(x)$}
      \label{fig:G_Lyapunov}
    \end{center}
  \end{center}  
\end{figure}
\begin{figure}[ht]
  \begin{center}
    \includegraphics[width=80mm]{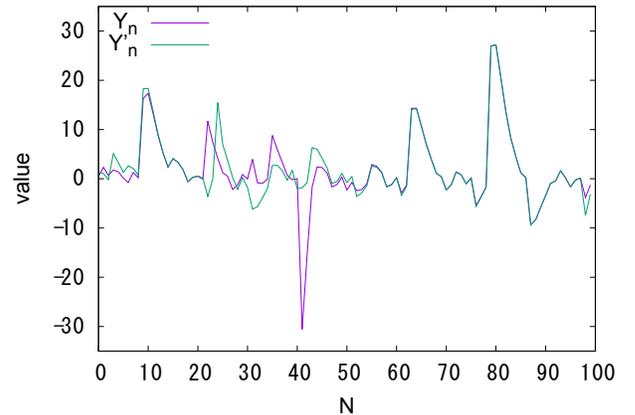}
    \begin{center}
     \caption{An example of a chaotic synchronization}
      \label{fig:synchronize}
    \end{center}
  \end{center}
\end{figure}
\begin{figure}[ht]
  \begin{center}
     \includegraphics[width=60mm]{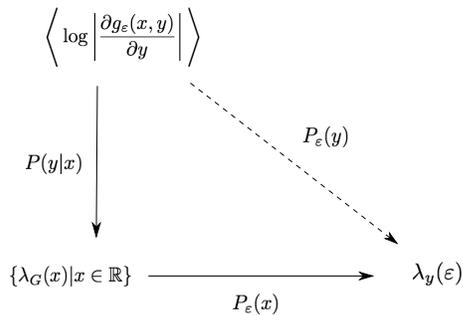}
    \begin{center}
     \caption{the new relation for CLE}
      \label{fig:new}
    \end{center}
  \end{center}
\end{figure}

Chaotic signal induced synchronization so called {\it chaotic
synchronization} is known to occur robustly in various physical models
\cite{Fujisaka_Yamada_1983,Peroca_Carroll_1990,Kocarev_Parlitz_1995,Kocarev_Parlitz_1996,Fujisaka_1997}.
  {\it The conditional Lyapunov exponents} (CLEs) are used for detecting
  chaotic synchronization and these play an important role in an
  indication of chaotic synchronization. However, most previous studies
  on chaotic synchronization obtains CLEs numerically, and none of the
  studies have proven chaotic synchronization for a given model except
  the numerical simulations while chaotic synchronizaton have been
  experimentally observed \cite{sunada2014,umeno_2005}. 
In this Letter, we show an analytically solvable model of chaotic
synchronization where the CLEs are exactly obtained in terms of the
coupling strength by the ergodicity of
{\it solvable chaos} \cite{umeno_1998,umeno_1997} used in our model. In addition, a new genaral relation of
CLE is established, which provide a new interpretation of a
 {\it common noise-induced synchronization} mechanism \cite{Teramae_Tanaka_2004}.\\
Here, we consider a class of dynamical systems with a
chaotic map $ f $ as follows
\begin{equation}\label{eq:system1}
  \begin{cases}
    X_{n+1} &= f(X_n)\\
    Y_{n+1} &= f(Y_n)+\varepsilon X_{n}\equiv g_{\varepsilon}(X_n,Y_n).
  \end{cases}
\end{equation}
 The dynamical system \eqref{eq:system1} can be a model for unidirectionally coupled
 system. Here, the system $f$ shows chaotic behavior while $\varepsilon$
 represents the strength of coupling connection.\\
We choose $f$ as a solvable chaotic map \cite{umeno_1998}
as follows
\begin{equation}\label{eq:f}
  X_{n+1}=\frac{1}{2}\left(X_{n}-\frac{1}{X_{n}}\right)\equiv f(X_{n}).
\end{equation}
The function $f$ is one of the generalized Boolean transformations, and is based
on the duplication fomula of a cotangent function.
In addition, the function $f$ satisfies $-\cot{2\theta}=f[-\cot{\theta}]$
\cite{umeno_1998}，and $X_n$ has an analytically solvable solution
$X_{n}=-\cot((\pi/2)2^n\theta_{0})$ with $X_0$ being an initial point.
The mapping $f$ is a two-to-one mapping, and the ergodic invariant measure
$\mu({\rm d}x)=\rho(x){\rm d}x$ of
dynamical system \eqref{eq:f} satisfies the probability
preservation relation (The Perron-Frobenius equation \ [PF
equation]), given by
\[
  \rho(z)=\sum_{x=f^{-1}(z)}\rho(x)\left|\frac{{\rm
 d}x}{{\rm d}z}\right|.
\]
We note that
 the standard Cauchy distribution, satisfying the PF
 equation\cite{umeno_1998}, and which shows that $f$ preserves the
 standard Cauchy Law.
\[
 \rho(x){\rm d}x=\frac{1}{\pi(x^2+1)}{\rm d}x \hspace{3mm} (\equiv
 {\rm C}(x;0,1){\rm d}x)
\]
In addition, the Lyapunov exponent of the dynamical system \eqref{eq:f}
is analytically calculated by using the invariant measure, by the ergodic equality.
\begin{eqnarray*}
  \lambda_{f} & = & \lim_{N \to \infty}\frac{1}{N}\sum_{n=0}^{N-1}\log\left|\frac{{\rm d} f(x)}{{\rm d}x}\right|_{x=X_n} \label{eq:lya_teigi}\\
 &=&\int_\mathbb{R}\frac{1}{\pi(x^2+1)}\log\left|\frac{1}{2}\left(1+\frac{1}{x^2}\right)\right|{\rm d}x =\log{2} \notag
\end{eqnarray*}
Here, we say that a chaotic system is solvable if the ergodic
invariant measure of dynamical system can be analytically obtained.
We define a chaotic synchronization as
follows.
When we give the same initial points $X_0$ and the different initial points
$Y_0$ and $Y'_0$ in two unieque dynamical systems, a chaotic synchronization
occurs for unidirectionally coupled dynamical systems \eqref{eq:system1} if the
condition \eqref{eq:chaos_synchronization} is satisfied for any $Y_0 \neq Y'_0$ almost everywhere.
\begin{equation}\label{eq:chaos_synchronization}
 \lim_{n \to \infty}|Y_n-Y'_n|=0 \ \ \text{a.e.}
\end{equation}
The CLE is an index of indicating
whether chaotic synchronization {\it actually} occur or not.
Given the variations in the mapping for variables $X_n$ and $Y_n$ in the dynamical system
\eqref{eq:system1}, we have the following variational equations 
for considering the sensitive dependency with respect to the initial
conditions. Here, $J_k$ is the Jacobian matrix satisfying the following linear relation.
\begin{eqnarray}\label{eq:henbun}
  \left(\begin{array}{@{\,}c@{\,}}
    \delta X_{n} \\
    \delta Y_{n}
  \end{array}
  \right)=\prod_{k=1}^{n}J_{k}
  \left(\begin{array}{@{\,}c@{\,}}
    \delta X_{0} \\
    \delta Y_{0}
  \end{array}\right)
\end{eqnarray}
When we define a matrix  $A_n$ as $\prod_{k=1}^{n}J_k=J_nJ_{n-1}\dots
J_{1}$, we can describe $A_n$ as follows.
\begin{eqnarray*}
  A_{n} & =\left(
  \begin{array}{@{\,}cc@{\,}}
   \displaystyle \prod_{k=1}^n\frac{\partial f(x)}{\partial x}
     \Biggl|_{x=X_{k-1}}\Biggr. & 0 \\
   \dots &
    \displaystyle
    \prod_{k=1}^n\frac{\partial
     g_\varepsilon(x,y)}{\partial y}\Biggl|_{\scriptsize{\begin{matrix}
					      x=X_{k-1},\\
					      y=Y_{k-1}
					      \end{matrix} }}\Biggr.
    \end{array}
  \right)
\end{eqnarray*}
Here, the limiting behaviour of the component (1,1) of $A_n$ has been found to be $2^n$ as $n \to \infty$ since the
Lyapunov exponent of the system \eqref{eq:f} is $\log 2$.
On the other hand, the CLE for variable $y$ in the dynamical system
\eqref{eq:system1} can be determined by the limiting behaviour of the component (2,2) of $A_n$.
In addition, we will show later that the CLE $\lambda_{y}$ in the dynamical system
\eqref{eq:system1} with the mapping \eqref{eq:f} is
analytically obtained in terms of the coupling strength $\varepsilon$ by
the ergodicity of the following long time average formula:
\begin{align}
\displaystyle \lambda_y&=\lim_{n \to
 \infty}\frac{1}{n}\log\prod_{k=1}^n\left|\frac{\partial
 g_\varepsilon(x,y)}{\partial y}\right|_{x=X_{k-1},y=Y_{k-1}}\notag\\
\displaystyle &=\lim_{n\to \infty}\frac{1}{n}\sum_{k=1}^{n}\log\left|\frac{\partial g_\varepsilon(x,y)}{\partial y}\right|_{x=X_{k-1},y=Y_{k-1}}.\label{eq:CLE_dif}
\end{align}
When CLE for the variable $y$ is {\it negative}, it can be seen from \eqref{eq:henbun} that the chaotic synchronization {\it
actually} occur satisfying \eqref{eq:chaos_synchronization}.
 Note that generally their CLEs are obtained by
numerical calculation (long-term average) according to the definition
\eqref{eq:CLE_dif} except our solvable case.
According to the previous studies \cite{umeno_1998,umeno_okubo_2016}, it is proved that the
dynamical system \eqref{eq:f} has ergodicity and the mixing property.
We consider the invariant measure of the coupled dynamical system
\eqref{eq:system1} with the solvable chaotic mapping \eqref{eq:f} in order to
show that this system is solvable.  Since the probability distribution of the variable $x$ is
already known to be the standard ${\rm Cauchy}$ distribution, it is
necessary to show that the
probability distribution of the variable $y$ converges to a certain probability
distribution $P_{\varepsilon}(y)$ analytically.
In order to show them, we consider the PF equation for the coupling relation \eqref{eq:system1}
with the mapping $f$ again and the nature of the
stable distribution in 
the superposition of {\rm Cauchy} distributions.\\
At first, we consider the PF equation for the
equation $z=f(x)$ where input variables $x$ follow a {\rm Cauchy}
distribution ${\rm C}(x;0,\gamma)$ with a scale parameter
$\gamma$. Here, C means that ${\rm C}(x;0,\gamma)=\frac{\gamma}{\pi(x^2+\gamma^2)}$. 
Since the $f$ is a two-to-one mapping, $P(z)$ satisfies Pf equation:
\begin{equation}
 P(z)|{\rm d}z| =P(x_1)|{\rm d}x_{1}|+P(x_2)|{\rm d}x_{2}|\label{eq:probable},
\end{equation}
where $x_1$ and $x_2$ $(x_1> x_2)$ are the solutions of the quadratic equation $z=f(x)$. They satisfy the following formula,:
\begin{equation*}
  \begin{cases}
   x_1+x_2=2z\\
  x_1x_2=-1.\\
   \end{cases} 
\end{equation*}
Considering these relationships, the probability distribution $P(z)$ is
obtained, as the following rescaled Cauchy distribution:
\begin{equation}\label{eq:P_z}
  P(z)={\rm C}(z;0,\gamma') \  \left(\gamma'=\frac{\gamma^2+1}{2\gamma}
 \  \text{\cite{shintani_2015,umeno_2016}}\right).
\end{equation}
Thus, we can show that the mapping $f$ preserves
{\rm Cauchy} distributions 
with a change of the scale
parameter as $\gamma \to \gamma'\ $
 for any scale parameters $\gamma$.
 Considering the invariance property of Cauchy distributions,
 and the superposition of Cauchy distributions,
 we get the recurrence
 equation \cite{shintani_2015} about the scale parameter $\gamma_n$ of the
{\rm Cauchy} distributions for $Y_n$ of \eqref{eq:system1}.
\begin{equation}\label{eq:gamma_f1}
 \gamma_{n+1}=\frac{\gamma_{n}^2+1}{2\gamma_{n}}+|\varepsilon|
\end{equation}
From this self-consistent recurrence equation in \eqref{eq:gamma_f1}, the scale parameter converges to the
fixed point of \eqref{eq:gamma_f1} $\gamma^*
(= |\varepsilon| + \sqrt{\varepsilon^2 + 1} (\ge 1))$. Thus, $P(y)$
converges to the probability distribution ${\rm C}(y:0,\gamma^*)$. 
Thus far, though we can set an initial distribution as a {\rm Cauchy}
distribution with any scale parameter $\gamma (>0)$,
these relations are satisfied for almost all the initial distributions
because the basis dynamical system \eqref{eq:f} has mixing property.
As above $P_{\varepsilon}(y)$ is expressed analytically, and we finally get the formula \eqref{eq:P(y)_sol}.
\begin{equation}
P_{\varepsilon}(y)={\rm C}(y;0,\gamma^*) \  \left(\gamma^*=|\varepsilon|+\sqrt{\varepsilon^2+1}\right)\label{eq:P(y)_sol}
\end{equation}
Secondly, because the probability distribution $P_{\varepsilon}(y)$ is obtained, the
CLE in \eqref{eq:CLE_dif}
can be expressed as a phase average by the ergodic theorem:
\begin{equation*}
 \lambda_{y}=\int_\mathbb{R}\frac{\gamma^*}{\pi(y^2+{\gamma^*}^2)}\log\left|\frac{1}{2}\left(1+\frac{1}{y^2}\right)\right|{\rm d}y.
\end{equation*}
This integration can be calculated analytically, and we get
the {\it analytical} CLE in terms of the strength of connection
$\varepsilon$ as
\begin{equation}\label{eq:CLE}
  \begin{split}
  \lambda_{y}(\varepsilon) &=2\log\left(\frac{\gamma^*+1}{\gamma^*}\right)-\log{2}\\
  &=2\log{(\sqrt{\varepsilon^2+1}-|\varepsilon|+1)}-\log{2}.
  \end{split}
\end{equation}
Then the threshold of synchronization $\varepsilon_c$ can also be obtained by the
solution of $\lambda_{y}(\varepsilon)=0$.
In the dynamical system \eqref{eq:system1}, two different initial conditions
$Y_0$ and $Y_0'$ synchronize if and only if the following condition is satisfied.
\begin{equation*}
  \lim_{n\to \infty}|Y_{n}-Y'_{n}| = 0 \hspace{4mm} \Leftrightarrow
 \hspace{4mm} |\varepsilon|>\varepsilon_c,\\
\end{equation*}
where $\varepsilon_{c}$ satisfies $\lambda_{y}(\varepsilon_c)=0$, giving
the critical coupling strength as $\varepsilon_c=1$.
FIG. \ref{fig:compare_cle} illustrates that this analytical CLE exactly
corresponds to the numerically obtained CLE by the fomula in \eqref{eq:CLE_dif}. 
As above, we show that the dynamical system \eqref{eq:system1} is
solvable when $f(x)$ is given by a solvable chaos mapping in
\eqref{eq:f} with the stable Cauchy law as an ergodic invariant measure, and derive the CLE and the threshold of
chaotic synchronization analytically. 
Furthermore, we can also get the CLE and the threshold for
another dynamical system \eqref{eq:system1} analytically. For example, we use another solvable chaos mapping $f$ as
$\displaystyle X_{n+1}=2X_n/\left(X_{n}^2-1\right)$ \cite{umeno_1998}
with the standard Cauchy law as the ergodic invariant measure, which is based on
the doubling formula
 of a tangent function, and consider the same unidirectional coupled
 system in \eqref{eq:system1}.
In this case, we get the following probability distribution $P_{\varepsilon}(y)$,
the CLE $\lambda_{y}(\varepsilon)$ and the threshold $\varepsilon_c$
satisfying $\lambda_{y}(\varepsilon)=0$ respectively:
\begin{equation*}
 P_{\varepsilon}(y) = {\rm C}(y;0,\gamma^*) 
\end{equation*}
 with $\gamma^{*}$ satisfying
 ${\gamma^*}^2-|\varepsilon|{\gamma^*}^2-\gamma^*-|\varepsilon|=0$, and
\begin{eqnarray*}
 \lambda_{y}(\varepsilon)&=&\int_\mathbb{R}\frac{\gamma^*}{\pi(y^2+{\gamma^*}^2)}\log\left|\frac{2(1+y^2)}{(1-y^2)^2}\right|{\rm
  d}y\\
&=&2\log\left(\frac{\gamma^*+1}{{\gamma^*}^2+1}\right)+\log{2},
\end{eqnarray*}
and
$\lim_{n\to \infty}|Y_{n}-Y'_{n}| = 0 \  \ \text{for}
\  \  |\varepsilon|>\varepsilon_c$, 
 where $\varepsilon_c$ satisfies $\lambda_{y}(\varepsilon_c)
 =0$, giving the correspoding critical coupling strength as $\varepsilon_c=0.78\cdots$.\\
Two series with different initial conditions $Y_0$ and
$Y_0'$ {\it synchronize} if and only if the coupling strength satisfies
 $|\varepsilon|>\varepsilon_c$. Note that this critical coupling
 strength $\varepsilon_c$ is {\it different} from that of the former
 case that $f$ is given by \eqref{eq:f}.\\
Next, we explain how the CLE relates to the coupling strength in a
genaral framework of common noise-induced synchronization.
The CLE can be considered one of the dynamical values ​​obtained as a result
from two-dimensional dynamical systems \eqref{eq:system1}, meaning an
orbital expansion rate according to a change of a dynamical variable.
Here, we decompose the analysis of the dynamical system
\eqref{eq:system1} into the following two stages.
We consider changes of the induced (skew-product) mapping at first, and then averege out these each orbital expansion rate.
We define the induced mapping $G_x(y)$ as follows:
\begin{equation*}
 G_x(y)=f(y)+x.
\end{equation*}
Here, the Lyapunov exponent of this mapping $G_x(y)$ depends on the
input value $x$ and consider as a skew product transformation. We analytically calculate the Lyapunov exponent of
$G_x(y)$ for each value $x$ by considering the distributions of $y$.
Note the probabirity distribution of $y$ can be considered as a
conditional probability $P(y|x)$ because $y$
depends on the value $x$.
Then, the method of deriving $P(y|x)$ is again from the probability
preservation relation as we obtain $P(y)$
in Eqs.\eqref{eq:probable}-\eqref{eq:P(y)_sol}.
Assume that input variables $y$ of $f$ that follow a {\rm Cauchy}
distribution ${\rm C}(y;c,\gamma)$. Then $P(z)$ is given by considering
the {\rm PF} equation for $z=f(y)$, as
\begin{eqnarray}\label{eq:c',gamma'}
\begin{split}
&P(z)=\frac{\gamma'}{\pi\left\{(z-c')^2+\gamma'^2\right\}}
 \hspace{3mm}=\ \ {\rm C}(z;c',\gamma'),\\
 &\text{where}\  \gamma'=\frac{\gamma(\gamma^2+c^2+1)}{2(\gamma^2+c^2)}
 , \text{and} \hspace{2mm}
 c'=\frac{c(\gamma^2+c^2-1)}{2(\gamma^2+c^2)}.
 \end{split}
\end{eqnarray}
This expression of $P(z)$ is more general than \eqref{eq:P_z} because it has
a change of the median as $c \to c'$.
When we consider the superpositon of distributions $f(y)+x$, we can
regard the constant value distribution $x$ in the mapping $G_x(y)$ as the delta
funcion $\delta(x)$.
Remark that the Delta function can be defined as a specific {\rm
Cauchy} distribution ${\rm C}(y;x,\eta)$ with the scale parameter $\eta$ $\to$ $0$.
Considering the dynamical system in this way,
 it is sufficient to consider a distribution of $f(y)+x$ within
 the class of Cauchy distributions.
 We get the following self-consistent recurrence equations about a
median $c$ and a scale parameter $\gamma$, as
\begin{equation}\label{eq:parameter'}\left\{
\begin{split}
 c_{n+1}&=\frac{c_n(\gamma_n^2+c_n^2-1)}{2(\gamma_n^2+c_n^2)}+x\\
 \gamma_{n+1}&=\frac{\gamma_n(\gamma_n^2+c_n^2+1)}{2(\gamma_n^2+c_n^2)}+\eta.
\end{split}\right.
\end{equation}
We get the convergence values $\hat{c}$ and $\hat{\gamma}$ as the {\it
stable} fixed point of \eqref{eq:parameter'} for $\eta \to
0$, where
\begin{equation*}
\begin{split}&\left\{
 \begin{split}
  \hat{c}&=x\\
  \hat{\gamma}&=\sqrt{1-x^2}
 \end{split}\ \ \ \ \ \ \  (\text{if}\ |x|<1),\right. \\
 &\left\{
 \begin{split}
  \hat{c}&=x+\text{sgn}(x)\sqrt{x^2-1}\\
  \hat{\gamma}&=0
 \end{split}\  (\text{if}\  |x|\ge 1),\right.
 \end{split}
\end{equation*}
and $\text{sgn}(x)=\left\{\begin{matrix}
			     1\ \ \text{if} \ x>0.\\
			     -1\ \text{if}\  x<0.
			    \end{matrix} \right.$\\ 
Thus, we finally obtain {\it the conditional probability distribution}
$P(y|x)$, as

\begin{equation}\label{eq:P(y|x)}
  P(y|x)=\left\{
  \begin{split}
   &{\rm C}(y;x,\sqrt{1-x^2}) &(|x|<1).\\
   & {\rm C}(y;x+{\rm sgn}(x)\sqrt{x^2-1}),0)  &(|x|\ge1).
    \end{split}\right.
\end{equation}
Finally, we estimate the Lyapunov exponent $\lambda_{G}(x)$ as the phase average,
and get it analytically by solving its integration, as 
\begin{eqnarray}
\lefteqn{  \lambda_{G}(x) = \Bigg\langle \log \left|\frac{{\rm d}
					    G_x(y)}{{\rm d} y}\right|
 \Bigg\rangle_{P(y|x)}}\notag\\
  &=&\int_\mathbb{R}P(y|x)\log{\left|\frac{1}{2}\left(1+\frac{1}{y^2}\right)\right|}{\rm
 d}y\notag\\
 & =& \begin{cases}
    \log(1+\sqrt{1-x^2}) \ \ \ \ (|x|<1)\\
       \log\left\{1+\frac{1}{(x+{\rm
     sgn}(x)\sqrt{x^2-1})^2}\right\}-\log2 \ (|x|\ge1),
 \end{cases}\label{eq:lambda_G}
\end{eqnarray}
 where $\langle \cdot \rangle $ represents the phase average. 
FIG. \ref{fig:G_Lyapunov} illustrates the relation between $x$ and $\lambda_G(x)$. 
As can be seen from FIG. \ref{fig:G_Lyapunov}, the Lyapunov exponent
of the dynamical system $G_x(Y_{n+1})=f(Y_n)+x$ {\it changes} depending
on the value $x$.
Note that for $|x|>1$, the mapping $G_x(y)$ become no longer a
chaotic map because of the negative Lyapunov exponent $\lambda_G(x)$, and the dynamical
system has a unique stable point for almost all the initial points.
Having $\lambda_G(x)$, then we calculate {\it the weighted average} of these Lyapunov
exponents according to the distribution of $x$.
Note that this value $x$ which is originally the variable $\varepsilon X_n$ that follow
${\rm C}(x;0,|\varepsilon|)$, as
$\frac{|\varepsilon|}{\pi(x^2+\varepsilon^2)}\equiv P_{\varepsilon}(x)$.\\
Thus we can calculate the averaged value that
equals to $\lambda_y$.
That is, the CLE can be given as {\it the weighted averages of the Lyapunov
exponents} $\lambda_G(x)$, that can be considered a new relation for CLE.
 \begin{eqnarray}
 \lambda_y &=&\Big\langle \lambda_G(x)\Big\rangle_{P(x)}\notag\\
 &=&
 \int_\mathbb{R}P_{\varepsilon}(x)\left(\int_\mathbb{R}P(y|x)\log\left|\frac{{\rm d}
 G_x(y)}{{\rm d} y}\right|{\rm d}y\right){\rm d}x
  \label{eq:CLE_solution}\\
  \bigl(\bigr.&=& 2\log{(\sqrt{\varepsilon^2+1}-|\varepsilon|+1)}-\log{2},
   \  \text{when}\  f \  \text{is} \ \text{\eqref{eq:f}}\bigl.\bigr)\notag
 \end{eqnarray}
FIG. \ref{fig:G_Lyapunov} illustrates that the mapping $G_x(y)$ is
{\it expansive} when $|x|<1$ while it is {\it attractive} when $|x|>1$.
For $P_{\varepsilon}(x)(=P(\varepsilon X_n))$, the increase in $|\varepsilon|$ is equivalent
to having an effect on widening the tails of the Cauchy distribution of the probability density
function $P(x)$.\\
Thus, the increase in $|\varepsilon|$ increases the {\it outer}
weights of $\lambda_G(x)$ in FIG. \ref{fig:G_Lyapunov} when we calculate
the CLE by the formula \eqref{eq:CLE_solution}.
Then, the CLE converts to {\it negative} when the weight of {\it attractive}
 mappings is relatively larger than the weight of {\it expansive} mappings.
In this way, the coupling strength $\varepsilon$ {\it relates} to CLE
and the origin of the common noise-induced synchronization.
Now remark that the CLE does not depend on whether $x$ is deterministic chaotic or
non-deterministic random noise, but purely depends on the distribution of $x$, namely
$P_{\varepsilon}(x)=P(\varepsilon X_n)$.
In fact, even when using the equation $\displaystyle
X_{n+1}=2X_n/\left(X_{n}^2-1\right)$ for the mapping $f$ with the same
standard Cauchy distribution ${\rm C}(x;0,1)$, the CLE of
the dynamical system \eqref{eq:system1} is confirmed to have the same
expression in \eqref{eq:CLE_solution}.\\
 Since the CLE represents the {\it phase average} of the orbit expansion rate,
 temporal orbit expansion rate can not accurately represent the
 synchronization behavior of the dynamical system. 
 In fact as shown in FIG. \ref{fig:synchronize}, we can confirm that an
 orbit expansion (de-synchronizaton)  
  can temporarily appear, even when
  the CLE is negative.
This fact can be understood from the fact that the effect
of the noise amplitude of the mapping in the dynamical system affects a
  temporal synchronization behavior, such as previously
 described in FIG. \ref{fig:G_Lyapunov}.
Finally, the background of the relation \eqref{eq:CLE_solution} is
explained by the concept of {\it marginalization} of probability.
Note that for the terms with the formula \eqref{eq:CLE_dif} and
\eqref{eq:CLE_solution}, the following relation is generally satisfied for
unidirectional coupled dynamical systems \eqref{eq:system1}.
 \begin{equation*}
  \frac{\partial g_\varepsilon(x,y)}{\partial y}=\frac{{\rm d} G_x(y)}{{\rm d} y}=\frac{{\rm d}f(y)}{{\rm d}y}.
 \end{equation*}
 Forthemore, the probability $P(y)$ is expressed in
 terms of the maginal probability density function $P(y|x)$ as
 \begin{equation*}
P(y)=\int_{\mathbb{R}}P(x)P(y|x){\rm d}x
 \end{equation*}
Thus, we obtain the relation \eqref{eq:CLE_solution} which gives a new
interpretation of the common noise-induced synchronization.
This relation can conceptually be depicted in FIG. \ref{fig:new} via the
marginal probability formula as:
\begin{eqnarray*}
 \lambda_y(\varepsilon) &=& \int_\mathbb{R}P_{\varepsilon}(y)\log \left|\frac{\partial
 g_\varepsilon(x,y)}{\partial y}\right|{\rm d}y\notag\\
 &=& \int_\mathbb{R}\left(\int_\mathbb{R}P_{\varepsilon}(x)P(y|x){\rm
 d}x\right)\log \left|\frac{{\rm d}
 f(y)}{{\rm d} y}\right|{\rm d}y\notag\\
 &=& \int_\mathbb{R}P_{\varepsilon}(x)\left(\int_\mathbb{R}P(y|x)\log\left|\frac{{\rm
								      d} G_x(y)}{{\rm d} y}\right|{\rm d}y\right){\rm d}x\\
 &=&\int_{\mathbb{R}}P_{\varepsilon}(x)\lambda_{G}(x){\rm d}x.
\end{eqnarray*}
 In conclusion, we have obtained a model of solvable chaotic synchronization,
 where we analytically have obtained the CLE, a threshold
 of chaos synchronization and an {\it exact} limiting distribution of the coupled the
 dynamical systems \eqref{eq:system1}.
Moreover, we also have obtained the new relation for the CLE by considering the
local analytical Lyapunov exponents of the induced (skew-product) transformation,
and have proven that a negative CLE causing the common noise-induced synchronization in
unidirectional coupled two-dimensional chaotic dynamical systems have been
consistent with the local Lyapunov exponent
$\lambda_G(x)$, which could have had a positive value causing temporally
expansive behabior.

\bibliography{basename of .bib file}

\begin{thebibliography}{10}
\bibitem{Fujisaka_Yamada_1983}
  H. Fujisaka and T. Yamada, Prog. Theor. Phys. \textbf{69}, 32 (1983)
\bibitem{Peroca_Carroll_1990}
 L. M. Pecora and T. L. Carroll, Phys. Rev. Lett. \textbf{64}, 821 (1990) 
\bibitem{Kocarev_Parlitz_1995}
  L. Kocarev and U. Parlitz, Phys. Rev. Lett. \textbf{74}, 5028 (1995)
\bibitem{Kocarev_Parlitz_1996}
  L. Kocarev and U. Parlitz, Phys. Rev. Lett. \textbf{76}, 1816 (1996)
\bibitem{Fujisaka_1997}
  H. Fujisaka, Prog. Theor. Phys. \textbf{98}, 775 (1997)
 \bibitem{sunada2014}
	 S. Sunada, K. Arai, K. Yoshimura, M. Adachi, Phys.
	  Rev. Lett. \textbf{112}, 204101 (2014)
 \bibitem{umeno_2005}
	A. Uchida and K. Umeno, National Congress of Theoretical
	and Applied Mechanics, \textbf{55}, 79 (2006) (In Japanese)
 \bibitem{umeno_1998}
	  K. Umeno, Phys. Rev. E. \textbf{58}, 2644 (1998)
  \bibitem{umeno_1997}
	 K. Umeno, Phys. Rev. E. \textbf{55}, 5280 (1997)
\bibitem{Teramae_Tanaka_2004}
  J. N. Teramae and D. Tanaka, Phys. Rev. Lett.  \textbf{93}, 204103-1 (2004)
 \bibitem{umeno_okubo_2016}
	 K. Umeno and K. Okubo,
	 Prog. Theor. Exp. Phys. 021A01 (2016)
 \bibitem{shintani_2015}
	 M. Shintani and K. Umeno, IEICE technical report, \textbf{115(178)}, 11
	 (2015) (In Japanese)
 \bibitem{umeno_2016}
	 K. Umeno, IEICE-NOLTA , 7, 14 (2016)

\end{thebibliography}

\end{document}